# As-rich GaAs(001) surfaces observed during As$_4$-irradition by scanning tunneling microscopy

Shiro Tsukamoto[a], Markus Pristovsek, Bradford G. Orr[b], Akihiro Ohtake, Gavin R. Bell[c], and Nobuyuki Koguchi

National Institute for Materials Science

1-2-1 Sengen, Tsukuba, Ibaraki 305-0047, JAPAN

## ABSTRACT

As-rich GaAs (001) surfaces are successfully observed during As$_4$-irradition by a system in which scanning tunneling microscopy (STM) and molecular beam epitaxy can be performed simultaneously. With a substrate temperature of 440 °C and an As$_4$ partial pressure of 2x10$^{-6}$ torr, reflection high energy electron diffraction patterns and reflectance anisotropy spectra confirm a c(4x4) As-stabilized surface. STM images clearly show alteration of the surface reconstructions while scanning. It is postulated that continual attachment / detachment of As molecules to and from the surface produces the observed dynamic behavior.

PACS: 6835Bs, 7960Bm, 8265Dp, 0779Cz, 6116Ch, 8764Dz, 8115Hi

a) e-mail: TSUKAMOTO.Shiro@nims.go.jp

b) present address: The University of Michigan, Ann Arbor, MI48109, U.S.A.

c) present address: University of Warwick, Coventry, CV4 7AL, U.K.

# As-rich GaAs(001) surfaces observed during As$_4$-irradition by scanning tunneling microscopy

Shiro Tsukamoto[a)], Markus Pristovsek, Bradford G. Orr[b)], Akihiro Ohtake, Gavin R. Bell[c)],

and Nobuyuki Koguchi

National Institute for Materials Science

1-2-1 Sengen, Tsukuba, Ibaraki 305-0047, JAPAN

The III-V semiconductor materials exhibit an enormous variety of surface reconstructions whose stability depends principally on temperature and group V overpressure or V:III flux ratio. For GaAs(001), growth by molecular beam epitaxy (MBE) generally takes place under As-rich conditions[1], and so the corresponding As-stable reconstructions are particularly important. Of these, the (2×4) and c(4×4) reconstructions have been widely studied[2,3]. Genuinely *in situ* surface sensitive probes, which can operate while the sample is growing in the MBE chamber, include reflection high-energy electron diffraction (RHEED)[4], total reflection angle X-ray spectroscopy (TRAXS)[5] and reflection anisotropy spectroscopy (RAS)[6]. However, scanning tunneling microscopy (STM), well known for its utility in problems of surface structure and MBE growth, has generally been applied with the STM remote from the MBE chamber[7-9]. In the case of As-rich GaAs(001), this is potentially problematic since it is thought that a complex exchange of As molecules to and from the surface takes place during growth[9,10]. Performing STM measurements away from the MBE environment inevitably prevents such dynamics from being observed. In this paper we report on the first observations of a GaAs(001)-c(4×4) surface under As$_4$ irradiation at elevated temperature, using an system in which both STM and MBE are incorporated in one chamber and can operate simultaneously[11,12].

Si-doped GaAs(001) 1º off <111>A (n = $2 \times 10^{18}$ cm$^{-3}$) substrates were prepared by standard solvent cleaning and etching procedures, and then loaded into the MBE chamber. After the oxide was removed at 600 $^o$C under an As$_4$ flux of 2x10$^{-5}$ torr, an undoped GaAs buffer layer of 1.0 μm was grown at 580 $^o$C with the following conditions: As$_4$/Ga flux ratio ~30, growth rate 1.0 μm/hour and background pressure of 2x10$^{-7}$ torr. This resulted in a smooth GaAs surface with single bilayer steps. The samples were then kept at 440$^o$C with As$_4$ partial pressure of 2x10$^{-6}$ torr in order to produce a c(4x4) phase surface as determined by reflection high energy electron diffraction (RHEED) and reflectance anisotropy spectra (RAS or RDS), shown in Fig.1. While maintaining this As$_4$ overpressure, STM images were obtained in constant current mode using sample biases of -3.0 V (filled states) and tunneling currents of 0.2 nA as shown in Fig.2. The STM images show steps, holes, and a few islands. During these sample conditions, no Ga flux and elevated temperature, there might be an initial adatom flux towards the steps, resulting in a very low concentration of the islands imaged[13].

In is interesting to note that when the scanning speeds were lower than 1.0x10$^4$ nm/s, the images became white, indicating an apparent constant surface height. This effect is probably due to the limited lifetime of As$_4$ on the surface. The adsorption lifetime of As$_4$ is less than 10$^{-5}$ s on GaAs(001) at 440 $^o$C[14]. The vertical resolution of the STM image was better than 0.3 nm, as judged by the observation of the surface reconstruction. With scan speeds less than 10$^4$ nm/sec, the residence time of the tip over a 0.3 nm region is greater then 10$^{-5}$ s. Therefore, if the scan speed is slow desorptions, and possibly adsorptions, of As$_4$ are likely as the tip is over an atomic site. This atomic motion results in very noisy images with an apparent constant height. However, if the scan speed was faster, so that the lifetime of As$_4$ was longer then the time it took the tip to travel over an atomic distance, the growth front of the substrate could be observed as shown in Fig.2. This scan rate dependent process is illustrated in Fig.3. This phenomenon was observed repeatedly and consistently, and does not occur in the absence of an incident As$_4$ flux.

Returning to the STM images of the GaAs surface, one immediately notices that the growth front was disordered. However, by RHEED observations, it showed a clear c(4x4) reconstruction, as shown in Fig.1, both before and after STM imaging. But, STM images showed no reconstruction on the surface. One reasonable explanation for this apparent contradiction is that there is an additional equilibrium partial layer of mobile As molecules. This layer is illustrated in Fig.3, on the c(4x4) surface reconstruction[15,16]. In order to understand the nature of this additional adlayer, we mapped the surface differences by subtracting two sequential images acquired one second apart. This timescale is short compared to complete transitions (about 100-1000s at 440 $^o$C[17]) but much longer than the molecular dynamics timescale discussed in the previous paragraph. Therefore, single adsorption/desorption events are likely to be observed. Shown in Fig 4 are two such images and their difference. The open circle marks the same location among these three images. In the adsorption map dark regions correspond to material that has been added to the surface between scans. Figure 5 shows the same information only for a larger scan size. Clearly the adsorption of adatoms occurs not only at step edges, but also on flat terraces.

In summary, we have use RHEED, RAS, and a system in which STM and MBE can be performed simultaneously, to examine the GaAs (001) surface with a substrate temperature of 440 $^o$C and an $As_4$ partial pressure of $2x10^{-6}$ torr. We find that STM images clearly show alteration of the surface reconstructions while scanning even though RHEED patterns and RAS spectra confirm a clear c(4x4) As-stabilized surface. It is postulated that detaching and attaching of $As_2$ or $As_4$ molecules may be the cause of the surface dynamics. In order to fully understand the dynamics of this phenomenon, further investigation is needed.

The authors wish to thank Dr.T.Ohno, NIMS, for his instructive discussions and helpful comments. This study was partially performed through Special Coordination Funds of the Ministry of Education, Culture, Sports, Science and Technology of the Japanese Government.

**Figure Captions**

Figure 1. (a) RHEED pattern and (b) RAS spectra of Ga-rich GaAs (001) surface with the substrate temperature of 440 $^{o}$C and the As$_4$ partial pressure of $2 \times 10^{-6}$ torr. For comparison, RAS spectra of the (2x4) and (4x2) surfaces are also shown.

Figure 2. STM image of GaAs (001) c(4x4) As-stabilized surface with the substrate temperature of 440 $^{o}$C and the As$_4$ partial pressure of $2 \times 10^{-6}$ torr. Constant current mode using sample biases of -3.5V (filled states), tunneling currents of 0.2 nA, and scanning speeds of $1.0 \times 10^{4}$ nm/sec. At the bottom is a series of STM images in the area of the open white rectangle on the top image.

Figure 3. A model of STM observation on the surface during As$_4$ irradiation.

Figure 4. Adsorption map of additional adatoms on c(4x4) surface by subtracting a image from previous one. The time difference between two images is one second.

Figure 5. Large area adsorption map of additional adatoms on c(4x4) surface.

**Figure 1 Tsukamoto et al.**

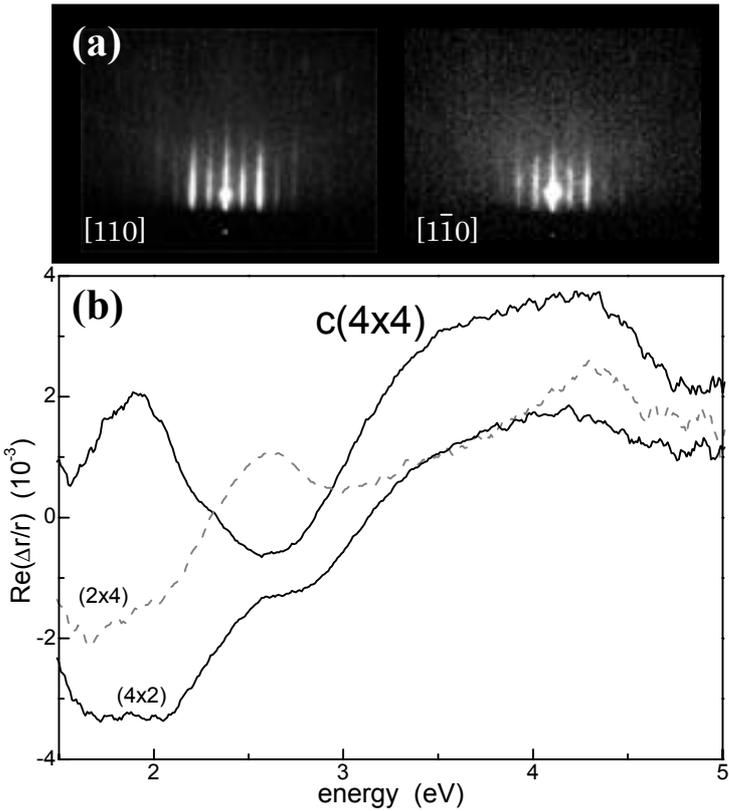

**Figure 2 Tsukamoto et al.**

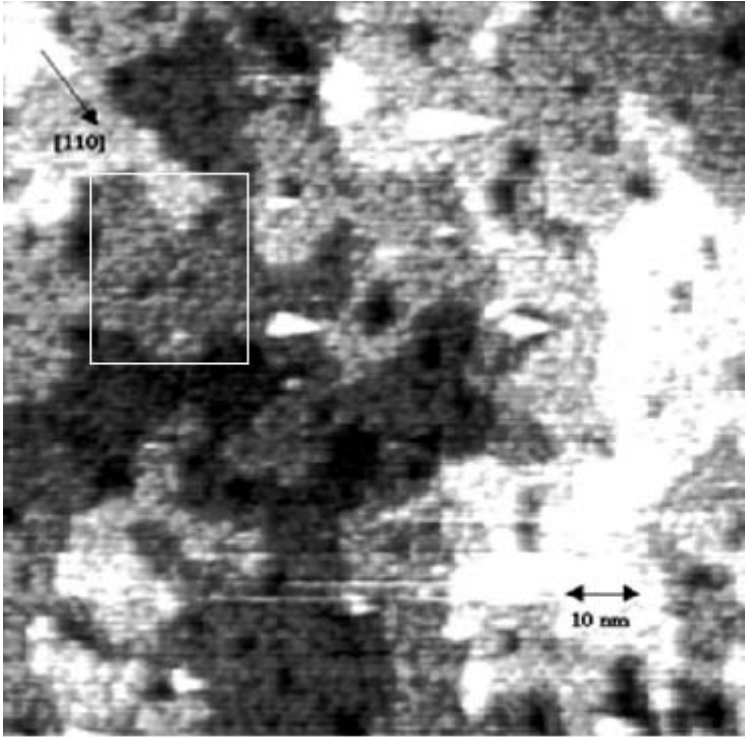
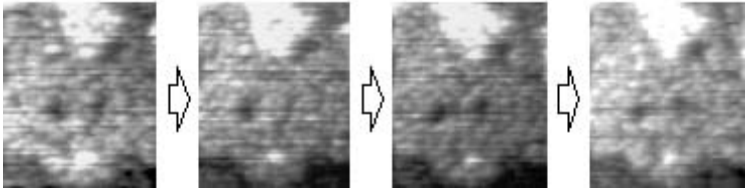

**Figure 3 Tsukamoto et al.**

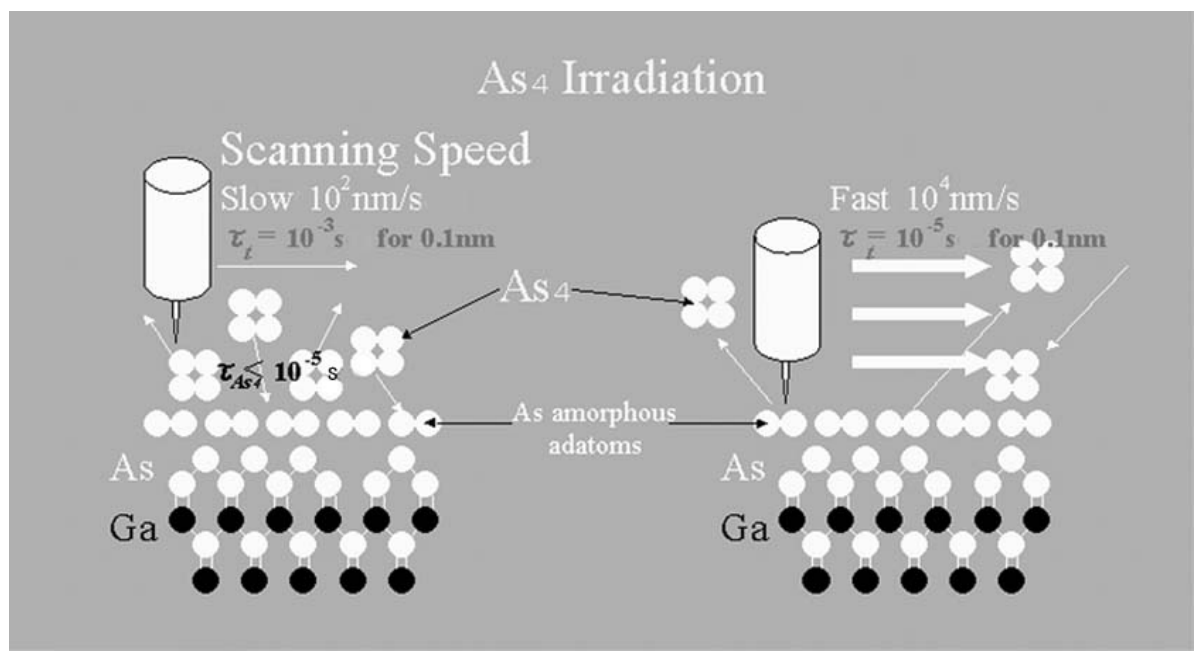

**Figure 4 Tsukamoto et al.**

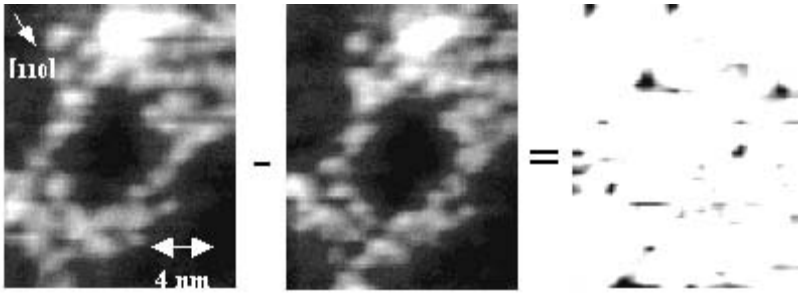

**Figure 5 Tsukamoto et al.**

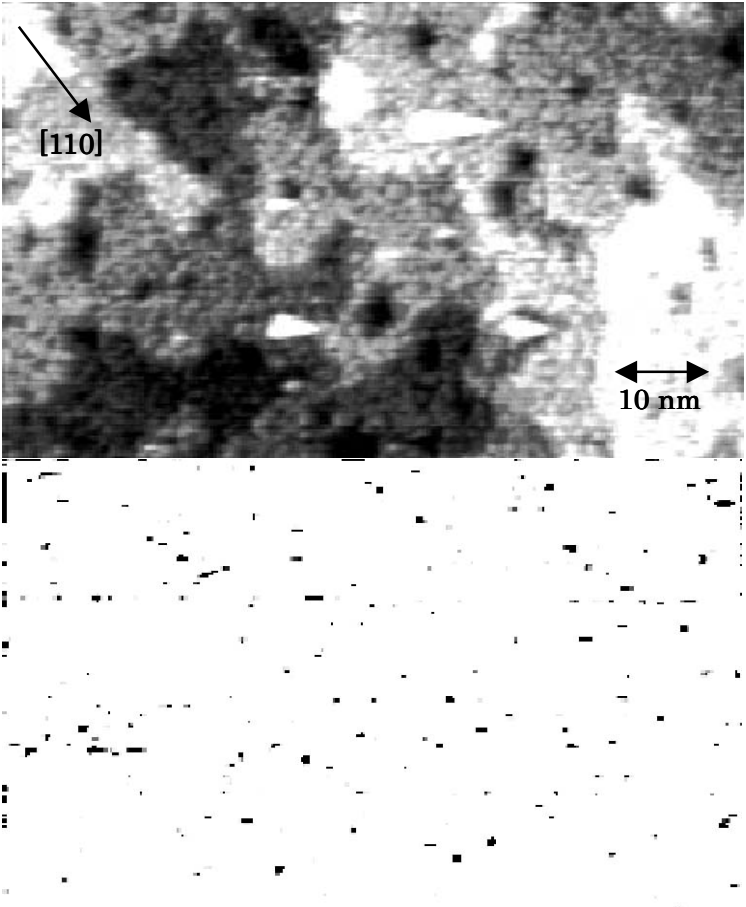